\documentclass[prl,twocolumn,showpacs,floatfix,amsmath,amssymb,superscriptaddress]{revtex4-1}
\usepackage{amsfonts}
\usepackage{stmaryrd}
\usepackage{bbm}
\usepackage{mathrsfs}
\usepackage{tipa}
\usepackage{amssymb}
\usepackage{txfonts}
\usepackage{graphicx}
\usepackage{dcolumn}
\usepackage{epstopdf}
\usepackage[colorlinks,linkcolor=blue,urlcolor=blue,citecolor=blue]{hyperref}
\usepackage{multirow}
\usepackage{subfigure}
\usepackage{url}
\usepackage{upgreek}
\usepackage[utf8]{inputenc}
\usepackage[english]{babel}

\begin{document}

\newcommand*{\cm}{cm$^{-1}$\,}
\newcommand*{\Tc}{T$_c$\,}

\title{Band-selective third-harmonic generation in superconducting MgB$_2$:\\ Possible evidence for Higgs amplitude mode in the dirty limit}

\author{Sergey Kovalev}
\affiliation{Institute of Radiation Physics, Helmholtz-Zentrum Dresden-Rossendorf, 01328 Dresden, Germany}

\author{Tao Dong}
\email{taodong@pku.edu.cn}
\affiliation{Institute of Physics, Johannes Gutenberg-University Mainz, 55128 Mainz, Germany}
\affiliation{International Center for Quantum Materials, School of Physics, Peking University, Beijing 100871, China}

\author{Li-Yu Shi}
\affiliation{International Center for Quantum Materials, School of Physics, Peking University, Beijing 100871, China}

\author{Chris Reinhoffer}
\affiliation{Institute of Physics II, University of Cologne, 50937 Cologne, Germany}

\author{Tie-Quan Xu}
\affiliation{Applied Superconductivity Center and State Key Laboratory for Mesoscopic Physics, School of Physics, Peking University, Beijing 100871, China}

\author{Hong-Zhang~Wang}
\affiliation{Applied Superconductivity Center and State Key Laboratory for Mesoscopic Physics, School of Physics, Peking University, Beijing 100871, China}

\author{Yue~Wang}
\affiliation{Applied Superconductivity Center and State Key Laboratory for Mesoscopic Physics, School of Physics, Peking University, Beijing 100871, China}

\author{Zi-Zhao Gan}
\affiliation{Applied Superconductivity Center and State Key Laboratory for Mesoscopic Physics, School of Physics, Peking University, Beijing 100871, China}

\author{Semyon Germanskiy}
\affiliation{Institute of Physics II, University of Cologne, 50937 Cologne, Germany}

\author{Jan-Christoph Deinert}
\author{Igor Ilyakov}
\affiliation{Institute of Radiation Physics, Helmholtz-Zentrum Dresden-Rossendorf, 01328 Dresden, Germany}

\author{Paul~H.~M.~van~Loosdrecht}
\affiliation{Institute of Physics II, University of Cologne, 50937 Cologne, Germany}

\author{Dong Wu}
\affiliation{International Center for Quantum Materials, School of Physics, Peking University, Beijing 100871, China}

\author{Nan-Lin Wang}
\affiliation{International Center for Quantum Materials, School of Physics, Peking University, Beijing 100871, China}
\affiliation{Beijing Academy of Quantum Information Sciences, Beijing 100913, China}

\author{Jure Demsar}
\affiliation{Institute of Physics, Johannes Gutenberg-University Mainz, 55128 Mainz, Germany}

\author{Zhe Wang}
\email{zhe.wang@tu-dortmund.de}
\affiliation{Department of Physics, TU Dortmund University, 44227 Dortmund, Germany}
\affiliation{Institute of Physics II, University of Cologne, 50937 Cologne, Germany}
\affiliation{Institute of Radiation Physics, Helmholtz-Zentrum Dresden-Rossendorf, 01328 Dresden, Germany}

\date{\today}

\begin{abstract}
We report on time-resolved linear and nonlinear terahertz spectroscopy of the two-band superconductor MgB$_2$ with the superconducting transition temperature $T_c \approx 36$~K. Third-harmonic generation (THG) is observed below $T_c$ by driving the system with intense narrowband THz pulses.
For the pump-pulse frequencies $f=0.3$, 0.4, and 0.5~THz, temperature-dependent evolution of the THG signals exhibits a resonance maximum at the temperatures with the resonance conditions $2f=2\Delta_\pi(T)$ fulfilled, for the dirty-limit superconducting gap $2\Delta_\pi$.
In contrast, for $f=0.6$ and 0.7~THz with $2f>2\Delta_\pi(T\rightarrow0)=1.03$~THz, the THG intensity increases monotonically with decreasing temperature. Moreover, for $2f<2\Delta_\pi(T\rightarrow0)$ the THG is found nearly isotropic with respect to the pump-pulse polarization. These results suggest a predominant contribution of the driven Higgs amplitude mode of the dirty-limit $\pi$-band superconducting gap, pointing to the importance of scattering for observation of the Higgs mode in superconductors.
\end{abstract}

\maketitle

Novel quantum phenomena, which may not be accessible to the linear spectroscopy in equilibrium states, can often be visualized in driven non-equilibrium states. A prominent example is the Higgs amplitude mode in a superconductor \cite{Anderson2,anderson1,PekkerVarma15,ShimanoTsuji20}.
With the $U(1)$ rotational symmetry spontaneously broken, a superconducting state is characterized by fluctuations of the amplitude and phase of the superconducting gap. The Higgs mode corresponding to the amplitude fluctuations is a scalar excitation without charge or electric/magnetic dipoles. As such, it cannot directly couple to external probes in the linear-response regime \cite{ShimanoTsuji20}.

Terahertz (THz) harmonic generation has been suggested to be an efficient probe to reveal Higgs mode in superconductors \cite{Matsunaga_NbN_Science,Higgs2,shimano_NbN_polarization,
JGWang_NPhoto19,Jigang2020,Chu20,Schwarz20}. Here, an intense narrowband THz pulse can drive a superconductor out of equilibrium, and trigger the coherent Higgs oscillations of the superconducting order parameter. The driven Higgs mode with twice the frequency of the pump pulse is further coupled to the pump pulse, leading to the third-harmonic generation (THG) \cite{ShimanoTsuji20}. Experimentally it is, however, not straightforward to unambiguously attribute the THG signal to the driven Higgs mode. In addition to a major enhancement of the THG signal below the superconducting transition temperature $T_c$, the Higgs mode excitation should give rise to a peak in the THG signal when twice the pump-pulse frequency matches the superconducting gap frequency. Although these features were observed in the \textit{s}-wave superconductor NbN \cite{Matsunaga_NbN_Science}, it was argued that the THG signal is more likely a result of charge density fluctuations due to the THz driven Cooper-pair breaking \cite{lara_CDF1}. The contribution of charge density fluctuations to the THG is, however, not always dominant, as pointed out by further theoretical studies \cite{phonon_retard_higgs,Schwarz20,haenel2021timeresolved}. Especially when correlation effects and/or impurity scattering are pronounced, the contribution due to the driven Higgs mode will dominate the THG signal. Moreover, the two competing scenarios suggested different dependences on the polarization of the driving THz field with respect to crystalline axis. NbN was found to exhibit an isotropic THG signal with respect to the polarization of the driving field, a signature supporting the assignment of the THG to the driven Higgs mode, whereas the THG from charge density fluctuations is expected to be anisotropic \cite{shimano_NbN_polarization}.

Recently, extensive theoretical studies of the THG in two-band superconductors addressed not only the Higgs amplitude fluctuations in individual bands, but also the so-called Leggett phase mode, resulting from the inter-band coupling \cite{MgB2theory3,MgB2theory2,MgB2theory1,Lara_polarization,Murotani19}.
Here, fluctuations of relative phase between the two-coupled order parameters, the Leggett mode, is also charge neutral and does not linearly couple to the electromagnetic field \cite{MgB2theory3}. Similar to the Higgs modes, the Leggett mode with frequency $\omega_L$ can be resonantly excited by a multicycle THz pulse with frequency $f$ satisfying the condition of $2f=\omega_L$, which is expected to be manifested by a resonance peak in the THG \cite{MgB2theory2,MgB2theory1,Lara_polarization,Murotani19}. Moreover, paramagnetic impurity scattering has been predicted to play a decisive role in the nonlinear response of the superconductors \cite{WuPRB17,Murotani19,impurity_Higgs,Seibold21,haenel2021timeresolved}.
While in the clean limit the THG was found to be dominated by pair breaking \cite{lara_CDF1,MgB2theory1}, the contribution of the Higgs mode can be significantly enhanced by impurity scattering. Thus, in dirty limit, the Higgs amplitude fluctuations are predominant than the charge density fluctuations, while the Leggett phase fluctuations in a two-band superconductor are much weaker and hardly detectable in a THG experiment \cite{Murotani19}.

Motivated by these theoretical advances, we performed THz spectroscopic measurements in the two-band superconductor MgB$_2$ in the linear- and nonlinear-response regimes. The linear response reveals the smaller gap $2\Delta_\pi$ to be in dirty superconductor limit. In contrast, no spectroscopic signature is evident in the vicinity of the larger gap $2\Delta_\sigma$ in the accessible spectral range, suggesting it may be in the clean limit. The THz third-harmonic generation is studied as a function of temperature for various driving frequencies and as a function of polarization with respect to the crystalline sample orientation. The temperature-dependent THG exhibits peaks at temperatures, where $2f$ matches $2\Delta_\pi$, whereas no anomalies are observed for $2f$ matching either the frequency of the larger gap or of the Leggett mode. These results point to a predominant contribution of the Higgs mode of the $\pi$-band in the dirty limit.

High-quality single-crystalline MgB$_2$ thin films with the \textit{c}-axis epitaxy were grown on MgO (111) substrate by using a hybrid physical-chemical vapor deposition method, and characterized by X-ray diffraction and charge transport measurements \cite{WangYue13,WangYue13JAP,Xi08}.
Optical measurements were carried out on a 13~nm thick MgB$_2$ films grown on $5\times5$ ~mm MgO substrate of 0.33 mm thickness, with $T_c \approx 36$~K \cite{WangYue13,WangYue13JAP}. Linear time-domain THz spectroscopy was performed using GaAs photoconductive antenna as THz source, driven by Ti:sapphire oscillator generating 35~fs pulses at 800 nm central frequency with the repetition rate of 80~MHz \cite{Shi18}. For the THG measurements, broadband THz pulses were generated via tilted pulse-front scheme utilizing a LiNbO$_3$ crystal \cite{Nelson07,Kovalev19}, driven by a Ti:sapphire amplifier generating 100~fs pulses at 800~nm with pulse energy of 1.5~mJ and repetition rate of 1~kHz. Spectral distribution of the generated THz pulses was optimized for around 0.7~THz central frequency. Narrow-band multicycle THz pulses were produced using bandpass filters with 20\% bandwidth.
The THz radiation was detected by electro-optic sampling at a ZnTe crystal.
Temperature dependent measurements were performed using helium-flow cryostats.



\begin{figure}
\includegraphics[width=8.5cm]{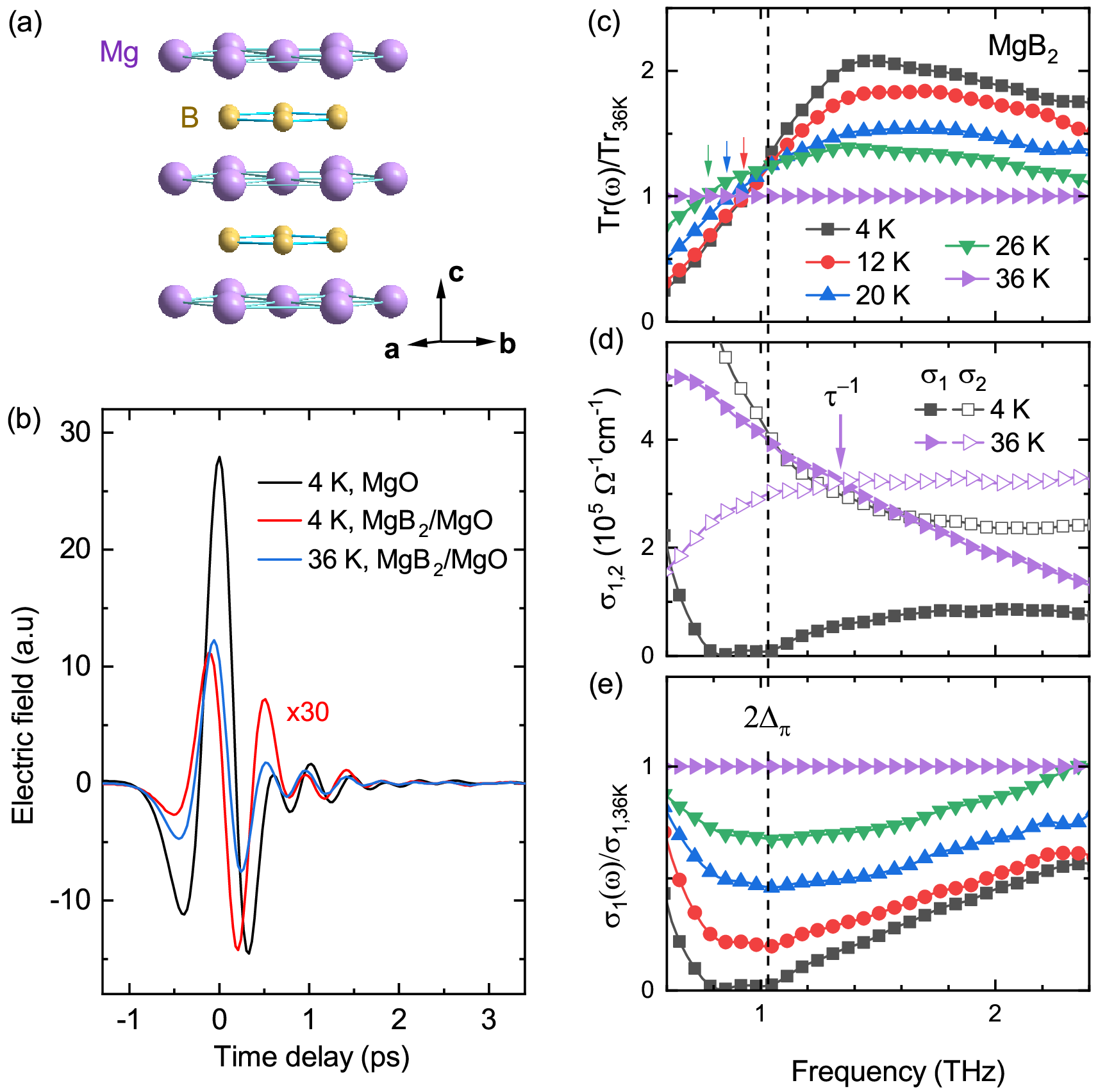}
\caption{\label{Fig_TDS}
(a) Layered crystallographic structure of MgB$_2$. THz electromagnetic waves propagated along the \textit{c}-axis.
(b) Transmitted THz electric field as a function of time delay, for MgB$_2$/MgO at 4 and 36~K and for the bare MgO substrate at 4~K.
The data for MgB$_2$/MgO at 4 and 36~K is magnified by a factor of 30 for clarity.
(c) Ratio of transmission for various temperatures with respect to 36~K. The arrows indicate the frequencies at which the ratio curves cross unity.
(d) Real $\sigma_1$ and imaginary $\sigma_2$ parts of optical conductivity at 36 and 4~K. The arrow indicates the crosspoint of the $\sigma_1$ and $\sigma_2$ curves at $\tau^{-1}=1.34$~THz for 36~K.
(e) Ratio of $\sigma_1$ for various temperatures with respect to 36~K. The dashed line indicates the excitation gap $2\Delta_\pi=1.03$~THz at 4~K.}
\end{figure}

The two-band BCS-type superconductor MgB$_2$ crystallizes in a hexagonal structure with the space group $P6/mmm$ [see Fig.~\ref{Fig_TDS}(a)] \citep{Akimitsu01,MazinPRL01a,Iavarone_STM_2002,MgB2_twogap_ARPES,Canfield03,XiRPP08}.
At 4.2~K two superconducting gaps of $2\Delta_\pi=1.1$ and $2\Delta_\sigma=3.6$~THz were found in the $\pi$- and $\sigma$-bands, respectively \cite{Iavarone_STM_2002}.
Figure~\ref{Fig_TDS}(b) shows THz electric field transients transmitted through the MgB$_2$/MgO sample at 4 and 36~K, compared to that through the reference bare MgO substrate at 4~K. By Fourier transformation of the time-domain signals, one can obtain transmission and phase shift in frequency domain, from which complex optical conductivity is derived.

Transmission ratio with respect to 36~K for various temperatures below $T_c$ is shown in Fig.~\ref{Fig_TDS}(c) as a function of frequency. For all the temperatures the transmission ratio increases first with decreasing frequency, and then drops continuously down to a value smaller than the unity.
The curves of the different temperatures below $T_c$ intersect with each other, forming an isosbestic point \cite{Vollhardt97,Greger13,Wang14}.
As indicated by the dashed line, the isosbestic point around 1.03~THz coincides with the onset of optical conductivity $\sigma_1$ at 4~K [see Fig.~\ref{Fig_TDS}(d)], which reflects formation of the superconducting gap in the $\pi$ band, in agreement with the reported results \cite{MgB2_THz_TDS,MgB2_twogap_ARPES,Iavarone_STM_2002,Canfield03}.

Figure~\ref{Fig_TDS}(d) shows the extracted complex optical conductivity in the normal state (36~K) and at 4 K, clearly demonstrating opening of the gap below 36~K. In the normal state, the data are consistent with the free carrier Drude response. The limited bandwidth prevents us from performing detailed analysis using two Drude components with reliable accuracy. Instead we estimate the scattering rate within the simple Drude scenario as the crossing point between $\sigma_1$ and $\sigma_2$. The extracted value $\tau^{-1}=1.34$~THz is indicated by the arrow in Fig.~\ref{Fig_TDS}(d). The estimated scattering rate in our sample is comparable to the values reported in Ref.~\cite{CarrHomes01,BBJin05}, but considerably lower than the values of 4.5 and 9 THz reported in Ref.~[\onlinecite{Pronin01}] and Ref.~[\onlinecite{MgB2_THz_TDS}],~respectively.

In the superconducting state the spectral weight of $\sigma_1$ at the finite frequency is depleted due to the opening of the superconducting gap as shown in Figs.~\ref{Fig_TDS}(d) and (e). At 4~K, $\sigma_1$ drops almost to zero below the superconducting gap at $2\Delta_\pi=1.03$~THz $\sim 4.26$~meV, while $\sigma_2$ exhibits a $1/\omega$ behavior.
We also note the upturn in $\sigma_1$ at frequencies below $\thicksim$ 0.7 THz, which is consistently observed in MgB$_2$ at low frequencies \cite{Pronin01,MgB2_THz_TDS}. This could be caused by an additional absorption of unpaired carriers, which seems to be intrinsic in nature \cite{Pronin01}.

The observed superconducting gap and temperature dependence of the optical conductivity are comparable with that reported for thin-film samples in Ref.~\cite{Pronin01,MgB2_THz_TDS}.
Quantitatively, broad-band normal state optical response was modelled also using two Drude terms, from which different scattering rates for the $\sigma$- and $\pi$-bands were derived \cite{vdMPRB06,Ortolani08}. Similar to the case in Ref. \cite{CarrHomes01,BBJin05}, our data suggest that $2\Delta_{\pi}<\tau^{-1}<2\Delta_{\sigma}$. Note that the scattering rates extracted in literature are generally higher in the $\pi$-band than in the $\sigma$-band \cite{vdMPRB06,Ortolani08}. Thus, our extracted value likely corresponds to the scattering rate of the $\sigma$-band. This suggests that superconductivity can be considered to be in the dirty limit for the $\pi$-band and in the clean limit for the $\sigma$-band. The contrast between the two bands seems crucial for the understanding of the nonlinear response, as presented in the following.

\begin{figure}
\includegraphics[width=8.5cm]{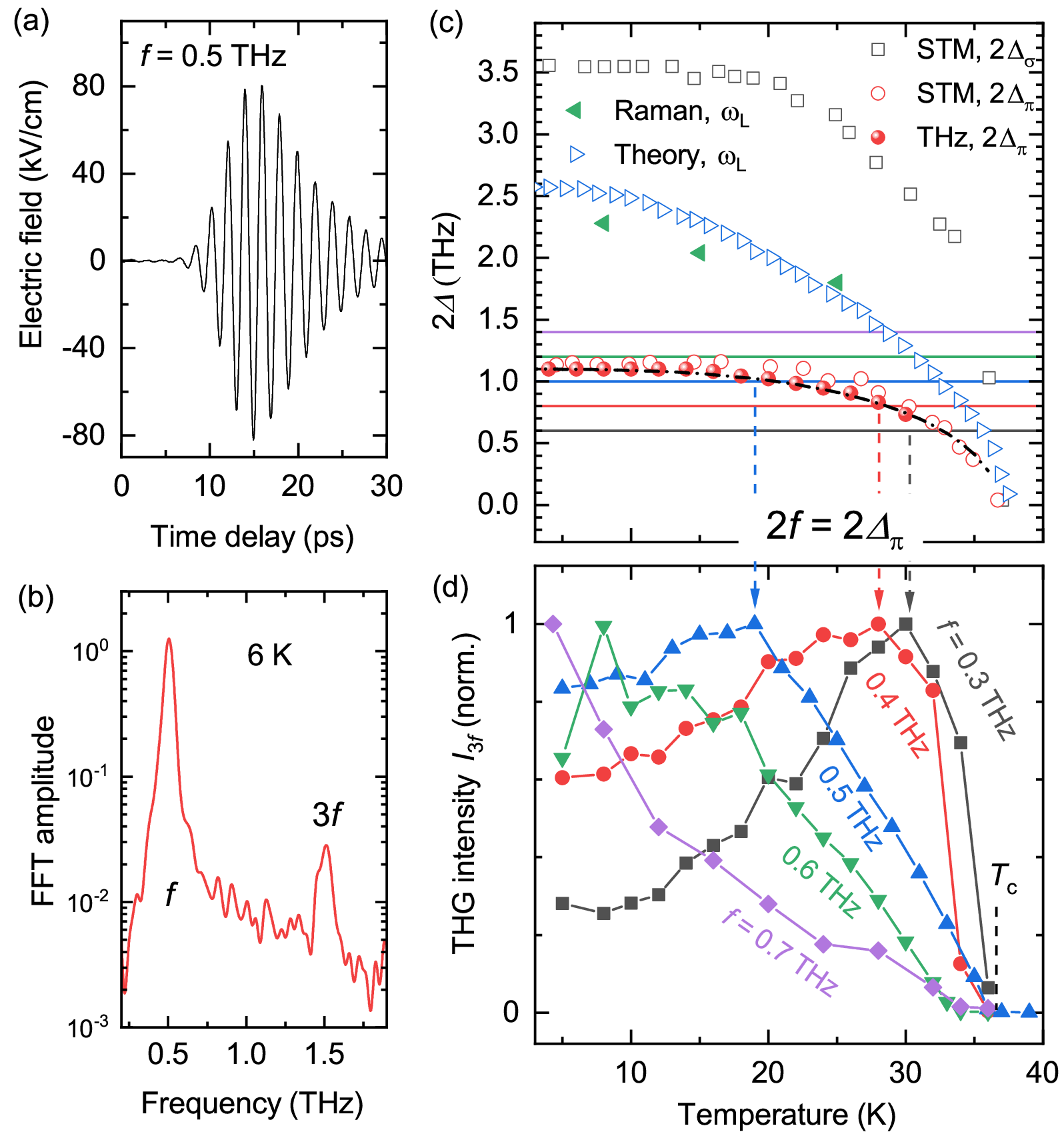}
\caption{\label{Fig_THG}
(a) Electric field of $f=0.5$~THz pump pulse in time domain.
(b) Spectrum of the generated radiation from the sample at 6~K. 
(c) Temperature dependence of superconducting gaps as determined by tunnelling spectroscopy (STM) for the $\sigma$ and $\pi$ bands (squares and circles) \cite{Iavarone_STM_2002}, and as estimated from Fig.~\ref{Fig_TDS}(c) (spheres).
Leggett mode versus temperature as observed in Raman spectroscopy (left-triangles) \citep{Leggett_Raman,Blumberg07Review} and obtained from theoretical calculations (right-triangles) \cite{Leggett_TPTR}.
The horizontal solid lines indicate $2f$ for $f= 0.3$, 0.4, 0.5, 0.6, and 0.7~THz, respectively.
(d) The corresponding normalized THG intensity $I_{3f}$ as a function of temperature, measured through two band-pass filters of $3f$. The vertical dashed lines indicate the peak positions in the temperature-dependent curves, where the corresponding resonance condition $2f=2\Delta_\pi$ is fulfilled simultaneously.
}
\end{figure}

Due to the low-frequency residual conductivity observed even at 4~K [Fig.~\ref{Fig_TDS}(e)], the temperature dependence of the superconducting gap $2\Delta_\pi(T)$ cannot be straightforwardly derived, e.g. by fitting the frequency-dependent optical conductivity according to the Mattis-Bardeen theory \cite{MgB2_THz_TDS}. To extract qualitatively the temperature dependence of the gap in the $\pi-$band, we approximate $2\Delta_\pi(T)$ by the point where the normalized transmission crosses the unity. As indicated by the arrows in Fig.~\ref{Fig_TDS}(c), the crossing point shifts continuously towards lower frequency with increasing temperature. By multiplying the extracted crossing point frequencies with the ratio of $2\Delta_\pi=1.03$~THz and the crossing point frequency at 4~K we obtain $2\Delta_\pi(T)$ as shown in Fig.~\ref{Fig_THG}(c) by the solid symbols. These results agree very well with the scanning-tunneling-spectroscopy data \citep{Iavarone_STM_2002}.

\begin{figure}
\includegraphics[width=8cm]{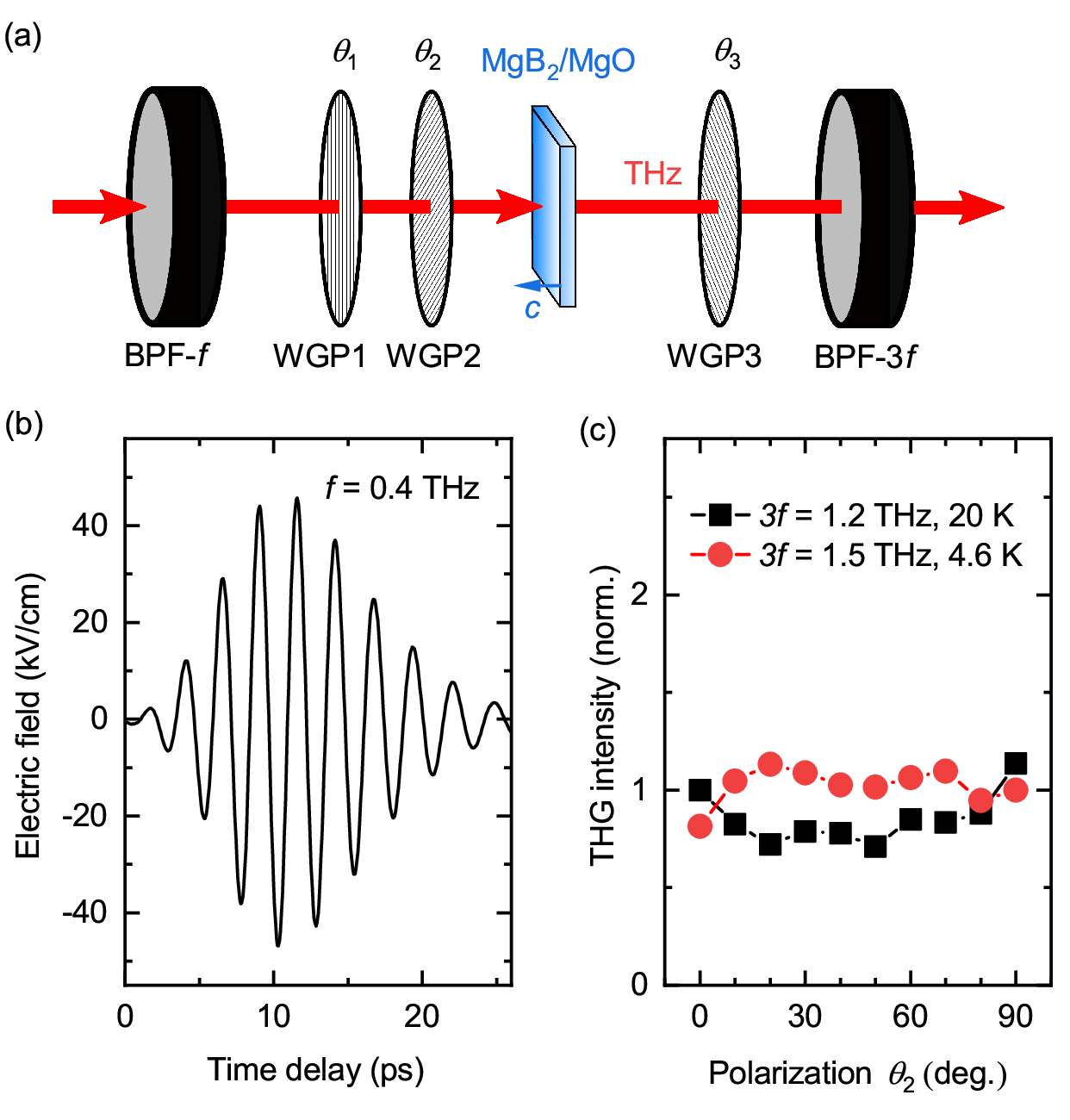}
\caption{\label{fig_polarz}
(a) Schematic illustration of the setup for polarization dependent THG measurements.
The THz electromagnetic waves propagate along the sample \textit{c}-axis.
BPF-\textit{f} and BPF-3\textit{f} denote bandpass filters for the fundamental frequency and the THG, respectively.
WGP's denote wird-grid polarizers.
(b) Waveform of the $f=0.4$~THz pump pulse for the polarization-dependent measurements.
(c) Normalized THG intensity as a function of the pump-pulse polarization, varied by tuning the angle $\theta_2$ of WGP2, which is obtained for $f=0.4$ and 0.5~THz at 20 and 4.6~K, respectively.
}
\end{figure}

We observed third-harmonic radiation in MgB$_2$ by pumping the system with multicycle THz pulses. Figure~\ref{Fig_THG}(a) displays the time-domain waveform of the $f=0.5$~THz pump pulse.
Driven by this pulse, the generated radiation of the sample was recorded with or without a $3f=1.5$~THz bandpass filter in time domain [see Fig.~\ref{fig_polarz}(a) for an illustrative sketch of the experimental setup].
Figure~\ref{Fig_THG}(b) shows the spectrum of the generated radiation without a $3f$-bandpass filter obtained via Fourier transformation of the time-domain signal at 6~K.

Temperature dependence of the obtained THG signal with a $3f$-bandpass filter is presented in Fig.~\ref{Fig_THG}(d). Starting from 5~K, the observed THG intensity for $f=0.5$~THz increases continuously with increasing temperature until about 19~K, which is followed by a continuous decrease at higher temperatures towards $T_c$. Above $T_c$, the THG signal is dramatically reduced. The enhancement of the third-harmonic generation below $T_c$ clearly reflects the nonlinear response of the superconducting state.
Moreover, the intensity maximum for $3f=1.5$~THz occurs right at a temperature where the resonance condition $2f = 2\Delta_\pi$ is fulfilled, which is indicated by the vertical dashed line in Figs.~\ref{Fig_THG}(c) and \ref{Fig_THG}(d).
However, such resonant behaviour is not observed in the vicinity of temperatures where the resonance condition would be fulfilled for the gap in the $\sigma$-band, or for the Leggett mode that was previously determined from Raman response \cite{Leggett_Raman,Blumberg07Review,Leggett_TPTR}, despite that such resonance peaks were predicted in the clean limit \cite{MgB2theory2,MgB2theory1}.
Although the observed THG signal here is dominated by the response of the $\pi$-band, 
one may still expect to reveal the Leggett mode and/or response of the $\sigma$-band via complementary pump-probe approaches, such as higher-frequency THz-pump broad-band THz probe \cite{ShimanoTsuji20,Leggett_TPTR}. A recent study shows that by using a pump pulse of 1.4~THz, which is evidently greater than the $\pi$-gap $2\Delta_\pi=1.03$~THz, it is possible to reveal signature of Leggett mode \cite{Leggett_TPTR}. For such a high-frequency pump pulse, response of the $\pi$-band should be primarily due to pair-breaking excitations [see Fig.~\ref{Fig_TDS}(e)].

To further study the temperature-dependent behavior,
we performed THG measurements for lower and higher pump-pulse frequencies, i.e. $f = 0.3$, 0.4, 0.6, and 0.7~THz.
As presented in Fig.~\ref{Fig_THG}(d), 
for $f=0.3$ and 0.4~THz, i.e. $2f<2\Delta_\pi(T \rightarrow 0)$, the temperature-dependent THG curves exhibit maxima at 30 and 28~K, respectively, matching the resonance condition $2f = 2\Delta_\pi$ (indicated by the arrows).
In contrast, for $2f>2\Delta_\pi(T \rightarrow 0)$, i.e. $f=0.6$ and 0.7 THz, the THG signal decreases monotonically with increasing temperature, without exhibiting a peak-like enhancement.
Although the $\sigma$-band located in the clean limit is in general expected to exhibit stronger THG signal due to charge density fluctuations, for none of the five frequencies we resolved a maximum corresponding to the resonance condition of the $\sigma$-band.
These observations suggest that the observed THG signal is unlikely being dominated by response from the $\sigma$-band (e.g. charge density fluctuations or Higgs amplitude fluctuations). Thus, we tend to ascribe the observed THG signals for $2f<2\Delta_\pi(T \rightarrow 0)$ to the Higgs amplitude fluctuations of the smaller $\pi$-gap in the dirty limit, which is in line with recent theoretical results \cite{Murotani19,haenel2021timeresolved}.

We further characterized the THG signal in the superconducting state as a function of pump-pulse polarization. 
As illustrated in Fig.~\ref{fig_polarz}(a), the polarization of the pump pulse is tuned by the wire-grid polarizer (WGP2) directly in front of the sample.
To assure a constant pump fluence at the sample position, another WGP1 is placed between WGP2 and the bandpass filter (BPF-\textit{f}) of the fundamental frequency.
After the sample, an additional wire-grid polarizer (WGP3) is inserted with a fixed polarization direction at $45^{\circ}$ with respect to the horizontal direction. With this polarizer, the electro-optic sampling at the ZnTe crystal (not shown) was optimized for detection. Before measuring the polarization dependence of the THG signals, the setup was calibrated by adjusting $\theta_1$ while setting $\theta_2$ for different polarizations, to ensure a constant pump fluence at the sample position.

Figure~\ref{fig_polarz}(b) shows waveform of the pump pulse of $f=0.4$~THz. The polarization-dependent THG intensity from the sample is presented in Fig.~\ref{fig_polarz}(c) for $3f=1.2$ and 1.5~THz at 20 and 4.6~K, respectively. The peak fields of the corresponding pump pulses were set at 47 and 54~kV/cm for $f=0.4$ and 0.5~THz, respectively. In the presented THG intensity, a factor, $\cos^2(\theta_2-\theta_3)$, due to the polarization difference between WGP2 and WGP3 has been taken into account. We can see that the THG intensity is only weakly dependent on the pump-pulse polarization for both the $f=0.4$ and the 0.5~THz pump pulses. It is worth noting that nearly polarization-independent THG is not necessarily the characteristic for the Higgs amplitude mode in general. On the one hand, isotropic THG was indeed theoretically predicted for the Higgs mode \cite{shimano_NbN_polarization,Schwarz20} and reported for the \textit{s}-wave superconductor NbN [\onlinecite{shimano_NbN_polarization}]. On the other hand, for multiband or unconventional superconductors (e.g. $d+s$ wave), theoretical studies showed that the THG due to the Higgs mode can also be polarization dependent \cite{Lara_polarization,Schwarz20,Seibold21}, whereas the charge density fluctuations may contribute to a nearly polarization independent THG signal \cite{Seibold21}. In this context, our experimental results appeal for material-specific theoretical analysis and place clear constraint on that.


To summarize, linear and nonlinear response of the superconducting state in the two-band superconductor MgB$_2$ has been studied by time-resolved terahertz spectroscopy. Third-harmonic generation was observed as a function of temperature for various pump-pulse frequencies. We found resonantly enhanced third-harmonic generation when twice the pump-pulse frequency matches the lower-energy gap $2\Delta_\pi$ that is in the dirty limit (i.e. $\tau^{-1}>2\Delta_\pi$). In contrast, such resonant behavior is absent for the higher-energy gap of the $\sigma$-band in the clean limit.
Since the THG for an \textit{s}-wave BCS superconductor in the clean limit should be dominated by charge density fluctuations, our results point to a dominant contribution to the THG from the Higgs amplitude fluctuations in the dirty-limit $\pi$-band, in agreement with theoretical analysis \cite{Murotani19,haenel2021timeresolved}. 
Nonetheless, complementary pump-probe approaches \cite{shimano_cuprate_800nm,Leggett_TPTR} are expected to address further aspects of the nonlinear response of the multiband superconductors, such as nonlinear response of a clean-limit band, polarization dependence, and effects of interband couplings  \cite{ShimanoTsuji20,haenel2021timeresolved,
MgB2theory2,MgB2theory1,MgB2theory3,Lara_polarization,
Murotani19,muller2021signatures}.

\begin{acknowledgments}
Z.W. gratefully acknowledges L. Benfatto, G. Blumberg, I. Eremin, F. Giorgianni, B. B. Jin, I.~I.~Mazin, R. Shimano, N. Tsuji, and J. Wang for illuminating discussions and/or communications. T.D. acknowledges financial support by the Alexander von Humboldt foundation. The work in Mainz was partially supported by the Deutsche Forschungsgemeinschaft
(DFG) under Grant No. TRR 173 268565370 (Project A05). The work in Beijing was supported by the National Science Foundation of China (No.11888101) and the National Key Research and Development Program of China (No. 2017YFA0302904). The work in Cologne was partially supported by the DFG  via  Project No. 277146847 — Collaborative Research Center 1238: Control and Dynamics of Quantum Materials (Subproject No. B05).

S.K. and T.D. contributed equally to this work.
\end{acknowledgments}

\bibliographystyle{apsrev4-1}
\bibliography{MgB2_bib}

\end{document}